\documentclass{aa}  
\usepackage{graphicx}  
\usepackage{txfonts}  
\begin{document}  
\headnote{Letter to the Editor} 
   \title{The calcium isotopic anomaly in magnetic CP stars
\thanks{  
Based on observations obtained at the European Southern Observatory, La Silla and  
Paranal, Chile (ESO programme Nos.\ 65.L-0316, 68.D-0254 and 266.D-5655)}  
}  
  
  
   \author{Charles R. Cowley  
          \inst{1}  
          \and  
          S. Hubrig\inst{2}\fnmsep  
          }  
  
   \offprints{C. Cowley}  
  
   \institute{Department of Astronomy, University of Michigan,  
              Ann Arbor, MI 48109-1090, USA\\  
              \email{cowley@umich.edu}  
         \and  
             European Southern Observatory,  
             Casilla 19001, Santiago 19, Chile\\  
             \email{shubrig@eso.org}  
             }  
  
   \date{Received ; accepted }

   \abstract{Chemically peculiar stars in the magnetic sequence  
can show the same isotopic anomaly in calcium previously discovered  
for mercury-manganese stars in the non-magnetic sequence.  In  
extreme cases, the dominant isotope is the exotic $^{48}$Ca.  
Measurements of \ion{Ca}{ii} lines arising from 3d-4p transitions reveal 
the anomaly by showing shifts up to 0.2~\AA{} for 
the extreme cases---too 
large to be measurement errors.  We report measurements of miscellaneous  
objects, including two metal-poor stars,
two apparently normal F-stars, an Am-star, and the N-star U Ant.
Demonstrable anomalies are apparent only for the Ap stars.  
The largest shifts are found in rapidly oscillating Ap stars and 
in one weakly magnetic Ap star, HD\,133792.  We note the possible
relevance of these shifts for the GAIA mission.
  
            }  
\titlerunning{$^{48}$Ca in magnetic stars}  
\authorrunning{Cowley and Hubrig}  
  
\maketitle  
%
  
\section{Introduction}  
\label{se:intro}  
Isotopic anomalies among the mercury-manganese (HgMn) chemically  
peculiar  
(CP) stars have been known since the early 1960's.  Bohlender et  
al. (2003), and Proffitt et al.\ (1999) give recent summaries and  
references.  
  
Castelli \& Hubrig (2004) described a new isotopic  
anomaly in HgMn stars that is revealed by wavelength shifts in the   
infrared triplet of \ion{Ca}{ii}. They 
discovered in the course of their abundance work on HR\,7143  
(HD\,175640) that the two available \ion{Ca}{ii} lines of the 
triplet were displaced 0.2\,\AA{} to the red of their expected  
positions.  The magnitude of the displacements for  
$\lambda\lambda$8498 and 8662 are just those expected  
if the calcium in the star were nearly pure $^{48}$Ca.  
The strongest line of the triplet, $\lambda$8542, was 
unavailable to Castelli \& Hubrig because of a gap  
in the coverage of their UVES spectra.  This line was  
(mostly) unavailable in the present work for the same  
reason.  

Castelli \& Hubrig noted the relevance of a large
isotopic shift in the \ion{Ca}{ii} infrared triplet for
the Gaia mission (cf.\ Katz et al.\ 2004).  While large shifts
may be confined to a few exotic objects, workers should
be aware that they can occur.

The relevant laboratory measurements were described by  
N\"{o}rtersh\"{a}user et al.\ (1998).  The  relatively  
large wavelength displacements are caused by mass-dependent,  
collective motions of the electrons and the atomic nucleus.  
These interactions cause the {\it specific mass shifts}  
(SMS).  The SMS are well known to be difficult to calculate  
accurately (Cowan 1981).  
  
Other lines of \ion{Ca}{ii} 
do not show these large shifts, which are due to the singular  
behavior of the 3d orbitals.   In \ion{Ca}{ii} the 3d--4p transitions 
make up the lines of Multiplet 2, 
$\lambda\lambda$8498, 8542, and 8662.  No transitions,  
other than those of the infrared triplet,  
involving 3d orbitals are available.  Transitions from 3d  
to 5p levels fall near 2130\,\AA{}.  Fortunately, the measured shifts 
are quite large (from  0.200\,\AA{} for $\lambda$8498 to 0.207\,\AA{}  
for $\lambda$8662) between  
$^{48}$\ion{Ca}{ii}  and  $^{40}$\ion{Ca}{ii}. 
  
\section{Stellar sample}  
  
The studied sample includes 19 magnetic chemically peculiar stars (Ap), 
two metal-poor halo stars, one Am star, the N-star U\,Ant, and Arcturus. 
Among Ap stars six are rapidly oscillating  Ap (roAp) stars which  
pulsate in high-overtone, low-degree, nonradial $p$-modes, with periods in the  
range 6--15~min (Kurtz 1990). The atmospheres of roAp stars 
are definitely abnormal showing remarkable ionization disequilibria of rare earths
elements (e.g.\ Ryabchikova et al.\ 2004). 
The basic data of our sample are presented in Table~\ref{tab:overview}. The columns   
indicate, in   
order, the HD number of the star, another identifier, 
and the spectral type, as it appears 
in the catalogue of Renson et al.\ (1991). 
The last two columns list the measured longitudinal field  or, if available, the surface   
magnetic field (in kG) and their source. 

All but one of the spectra have been obtained with the
VLT UV-Visual Echelle Spectrograph 
UVES at UT2. Most of the spectra of magnetic stars   
have been retrieved from  
the ESO UVES archive (ESO programme No.\ 68.D-0254). 
Spectra of one roAp star, HD\,24712, one 
Am star, HD\,27411, one N-star, HD\,91793, and two metal-poor stars, 
HD\,140283 and HD\,122563 were downloaded from the UVESPOP web 
site (Bagnulo et al.\ 2004). 
All spectra were observed with Dichroic standard settings covering the 
spectral range from 3030 to 10000\,\AA{} at the resolving power   
of $\lambda{}/\Delta{}\lambda{} \approx 0.8\times10^5$. 
They were reduced by 
the UVES pipeline Data Reduction Software (version 1.4.0),  
which is an evolved version of the ECHELLE context of MIDAS.  
One additional spectrum of HD\,101065 used in this study was obtained with 
the echelle spectrograph FEROS (Fiber Range Optical 
Spectrograph) on the 1.52\,m 
telescope at La Silla.  It covers the wavelength region 
between 3530 and 9220\,\AA{}, and has  
a nominal resolving power of 48000. 
To reduce the spectrum we used the standard MIDAS 
pipeline for FEROS. The FEROS spectum of HD\,101065 was used to confirm 
the shifts of the infrared \ion{Ca}{ii} triplet lines observed in the  
UVES spectrum of this star.  
\begin{table}  
\caption{Basic data of studied stars.}  
\label{tab:overview}  
\begin{tabular}{rllcc}  
\hline\noalign{\smallskip}  
\multicolumn{1}{c}{HD} &  
\multicolumn{1}{c}{Other} &  
\multicolumn{1}{c}{Sp.\ Type} &  
\multicolumn{1}{c}{$\left<{\cal B}_z\right>$/ $\left<{\cal B}_s\right>$} &  
\multicolumn{1}{c}{Ref.} \\
\hline\noalign{\smallskip}  
739 & $\theta$ Scl&F4V&\multicolumn{1}{c}{--} & $-$ \\  
965 & BD -00$^{\circ}$21 & A8 Sr& $-$/4.4& 1\\  
24712 & HR\,1217& A9 SrEuCr roAp & 1.2/$-$ & 2  \\  
27411 & HR\,1353& A3m&\multicolumn{1}{c}{--}&$-$\\  
47103 & BD -20$^{\circ}$1508& A SrEu & $-$/17.5& 1\\   
50169 & BD-01$^{\circ}$1414& A3 SrCrEu & $-$/4.7 & 1\\ 
60435 & CD-57$^{\circ}$1762& A3 SrEu roAp & $-$0.4/$-$& 5 \\ 
75445 & CD-38$^{\circ}$4907& A3 SrEu & $-$/3.0 & 1\\ 
91793 & HR\,4153 & N-star &\multicolumn{1}{c}{--}& $-$\\  
93507 & CD-67$^{\circ}$955 &A0 SiCr & $-$/7.2 & 1\\  
101065 & CD-46$^{\circ}$7232& F3 roAp& $-$1.0/$-$& 3 \\ 
116114 & BD-17$^{\circ}$3829& F0 SrCrEu & $-$/6.0 &1\\  
122970 & BD+06$^{\circ}$ 2827&F0 roAp & 0.2/$-$ & 4\\  
122563 & HR\,5270 &F8 IV halo star&\multicolumn{1}{c}{--}& $-$\\ 
124897 & HR\,5340 &K1.5 III Arcturus &\multicolumn{1}{c}{--}& $-$\\ 
133792 & HR\,5623& A0p SrCr & $-$/1.1& 5 \\ 
134214 & BD-13$^\circ$4081  & F2 SrEuCr roAp & $-$/3.1 & 1 \\ 
137909 & $\beta$\,CrB & A9 SrEuCr& $-$/5.5&1\\   
137949 & 33\,Lib & F0 SrEuCr roAp & $-$/4.7&1\\  
140283 & BD-10$^{\circ}$4149 & sdF3 halo star&\multicolumn{1}{c}{--}&$-$\\ 
146836 & HR\,6077& F6 III& $-$ \\ 
176232 & 10\,Aql & A6 Sr roAp & $-$/1.4 & 6\\  
187474 & HR\,7552 &A0 EuCrSi & $-$/5.0 &1\\  
188041 & HR\,7575& A6 SrCrEu & $-$/3.7&1\\  
216018 & BD-12$^{\circ}$6357 &A7 SrCrEu & $-$/5.6 &1\\  
217522 & CD-45$^{\circ}$14901& A5 SrEuCr roAp & $-$0.7/$-$& 3\\  
\hline\noalign{\smallskip}  
\end{tabular}  
(1) Mathys et al.\ (1997); (2) Leone \& Catanzaro (2004); 
(3) Hubrig et al.\ (2004a); (4) Hubrig et al.\ (2004b); 
(5) Mathys \& Hubrig, in preparation;  
(6) Leone, Vacca \& Stift (2003) \\ 
  
\end{table}  
\section{Wavelength determination and accuracy}  
  
ASCII files of the ESO spectra were interpolated to give a point
every 0.02\,\AA{}, and mildly filtered  
using a standard Brault-White (1971) algorithm.  Several methods were  
used to establish the zero point of the stellar atomic lines.  
Most commonly, a radial velocity scan was performed, using 
wavelength coincidence statistics (Cowley \& Hensberge 1981)
for the atomic
species expected to be strongly present.  The radial velocity  
of the star was taken to be that which gave the most highly  
significant results.  The method works very well. 
Uncertainties under a km/sec, and typically 
half a km/sec, result.

For the magnetic Ap stars, it can be difficult to obtain a good
radial velocity from the regions of the spectra that contained  
the two infrared \ion{Ca}{ii} lines.  There are relatively
few easily identifiable strong lines in this region, and the
Zeeman splitting increases quadratically with wavelength.
The magnetic null lines
\ion{Fe}{ii} $\lambda$7224 and \ion{Fe}{i} $\lambda$7389 were 
useful as a check in such instances, and in a few cases, 
the \ion{O}{i} triplet $\lambda\lambda$7772, 7774, and 7775 
could be used.  In some cases we used a nearby
region (e.g. $\lambda\lambda$5817--6834) acquired on the same  
night and at nearly the same time.
  
Both \ion{Ca}{ii} lines are found near gaps in echelle orders.  The 
problem is particularly severe for the $\lambda$8668 line, but  
may also be significant for $\lambda$8498.  We give two 
arguments that the coherence of the wavelength scale is sufficient 
to provide wavelengths that are accurate for the present 
purpose.  First, we obtain
reasonable wavelengths for normal stars.  Second, in the 
instances where we claim the stars show evidence of $^{48}$Ca, 
both \ion{Ca}{ii} lines are shifted by appropriate amounts 
(see Fig.~\ref{fig:one} described below).
We claim an overall wavelength
accuracy in the range 0.03 to 0.04\,\AA{}; one or two 
measurements could be in error by as much as 0.05\,\AA{}. 
 
Solar wavelengths were  
simply taken from the Rowland Tables (Moore, Minnaert \&  
Houtgast 1966).  For Arcturus, we adopted the wavelength  
calibration of Hinkle, Wallace \& Livingston (1995).  
The sunspot spectrum is from Wallace et al.\ (1998).

\section{Results}  
  
Table~\ref{tab:wl} gives the wavelength measurements for 
our sample (including the sun, Arcturus, and a 
sunspot).  
We also give the wavelength from the NIST  
site (Sansonetti \& Martin 2003).  
%
   \begin{figure*} 
   \centering  
     \includegraphics[width=0.6\textwidth,angle=270]{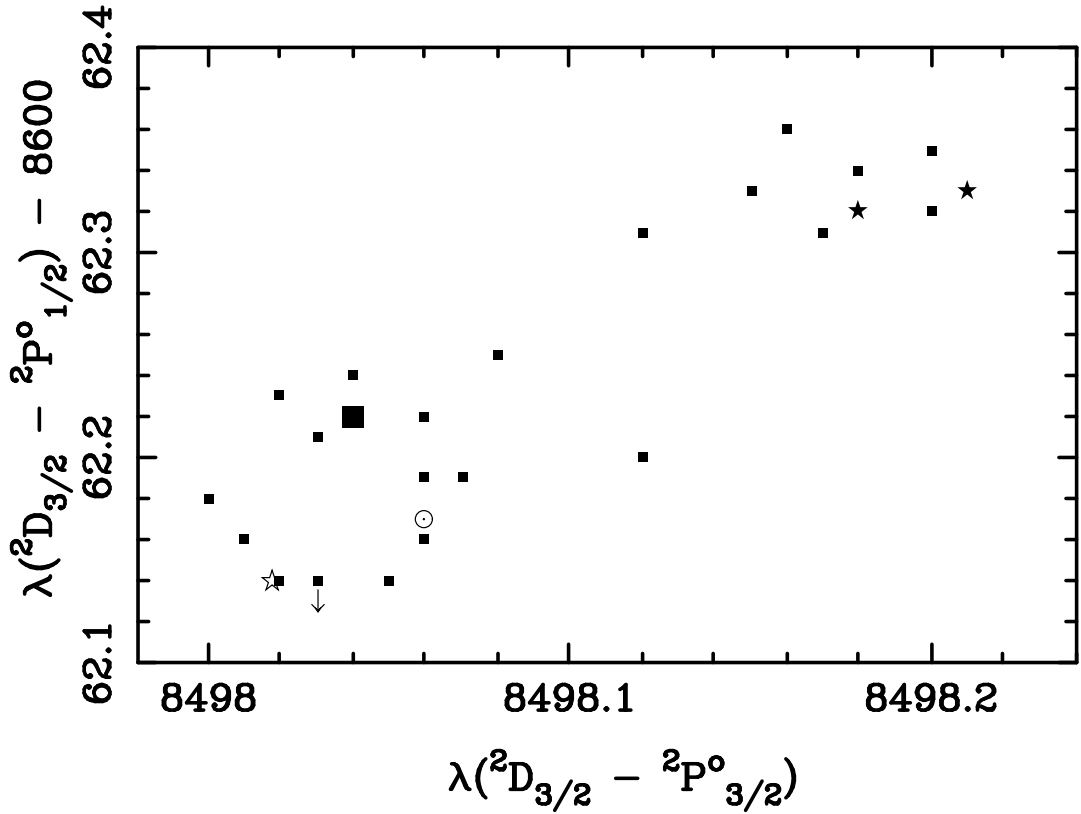} 
      \caption{Wavelengths for the \ion{Ca}{ii} $\lambda$8662-line 
              vs. those of the $\lambda$8498-line.  The open star  
              is the laboratory position (NIST).  The sun symbol  
              marks the photospheric wavelengths.  
              Stars showing the $^{48}$Ca anomaly are in 
              the upper-right portion of the diagram.  There  
              are two points (filled stars) for HD 101065.
         The large square
         is for HD 27411 and HD 187474 whose points overlap.  The two
         halo stars, HD 140283 and HD 122563 are not plotted because 
         the $\lambda$8662 line was not available.  The sunspot 
         was also not plotted because of the complexity of the 
         $\lambda$8662 profile. 
         } 
         \label{fig:one}  
   \end{figure*} 
%

The data of Table~\ref{tab:wl}  
are plotted in Fig.~\ref{fig:one}.  The stars generally fall 
into two groups. 
One group has points typically within a 0.04\,\AA{} 
radius including the sun's position.  The scatter may be 
reasonably attributed to various calibration and  
measurement uncertainties, blending, and macroturbulence. 
 
There is a net positive displacement of the 
cluster in the lower left from the laboratory 
position, marked by an open star.  Both wavelengths are
affected in the same sense.  The solar wavelength of the 
third member of the \ion{Ca}{ii} triplet, $\lambda$8542 is also
slightly displaced, positively, from the laboratory 
position.  The possibility that this is due to an 
isotopic variation of solar and terrestrial material 
is intriguing.  It is, of course, not the first such 
example.  In any event, the stellar positions in the 
lower left grouping of points in Fig.~\ref{fig:one} 
show the same trend.

A second group consists of seven stars; measurements for 
Przybylski's star are from two spectrographs.  Six 
objects are rapidly-oscillating  Ap stars.  The six are grouped together 
at the end of Table~\ref{tab:wl}. 
One of the stars in this group, HR 5623 (HD 133972) is 
not a roAp, and one roAp, 33 Lib, does not show the shift. 
Two objects thought to be related to the roAp stars because 
of their core-wing anomaly (Cowley et al.\ 2001 ),
HD 965 and HR 7575 (HD 188041) 
also do not 
have the $^{48}$Ca shifts. 
 
Three points fall between the two groups.  It is not yet 
clear to what extent they may be genuine intermediate 
cases or whether they 
are due to large errors.  We admit the latter might be as 
large as 0.05~\AA.

  
   \begin{table}  
  
     \caption{Observed positions of \ion{Ca}{ii} lines.} 
         \label{tab:wl}  
   \begin{tabular}{l l l} \hline\hline 
\multicolumn{1}{c}{8498}    &\multicolumn{1}{c}{8662}         &Star\\  
        \hline  
  
8498.018&    8662.140 & NIST                             \\  
8498.06 &    8662.17  & sun                         \\
8498.09 &    8662.22: & sun sunspot                      \\  
8498.02 &    8662.14  & HD 124897, Arcturus                         \\  
8498.07 &    8662.19  & HD 739, $\theta$ Scl, HR 35, F4 V\\  
8498.06 &    8662.19  & HD 146836, HR 6077, F6 III \\ 
8498.04 &    8662.22  & HD 27411, HR 1353, Am            \\ 
8498.05 &    8662.14  & HD 91793, U Ant, HR 4153, N-star  \\ 
8498.04 &    8542.11  & HD 140283, halo dwarf             \\ 
8498.05 &    ---\footnotemark       & HD 122563, halo giant             \\ 
        \hline 
\multicolumn{3}{c}{Ap stars} \\ 
        \hline 
8498.04 &    8662.24  & HD 965         \\ 
8498.00 &    8662.18  & HD 47103     \\ 
8498.03 &    8662.21  & HD 50169     \\ 
8498.08 &    8662.25  & HD 60435               \\ 
8498.12 &    8663.31  & HD 75445                   \\ 
8498.03 & $\le$8662.14& HD 93507  \\ 
8498.06 &    8662.16  & HD 116114                \\ 
8498.20 &    8662.32  & HD 133792, HR 5623 \\ 
8498.12 &    8662.20  & HD 137909, $\beta$ CrB  \\ 
8498.02 &    8662.23  & HD 137949, 33 Lib       \\ 
8498.06 &    8662.22  & HD 188041, HR 7575               \\ 
8498.04 &    8662.22  & HD 187474, HR 7552                \\ 
8498.01 &    8662.16  & HD 216018 \\ 
        \hline 
\multicolumn{3}{c}{roAp stars} \\ 
        \hline 
8498.17 &    8662.31  & HD 24712, HR 1217         \\ 
8498.21 &    8662.33  & HD 101065 UVES      \\ 
8498.18 &    8662.32  & HD 101065 FEROS     \\ 
8498.15 &    8662.33  & HD 122970           \\ 
8498.20 &    8662.34  & HD 134214           \\ 
8498.16 &    8662.36  & HD 176232, 10 Aql  \\ 
8498.20 &    8662.35  & HD 217522           \\  \hline 
 \end{tabular} 
\footnotesize{$^1$NIST position is 8542.09;}
\footnotesize{$\lambda$8662 region not observed.}
 \end{table}  
  
\begin{figure}  
   \centering  
   \includegraphics[width=0.45\textwidth]{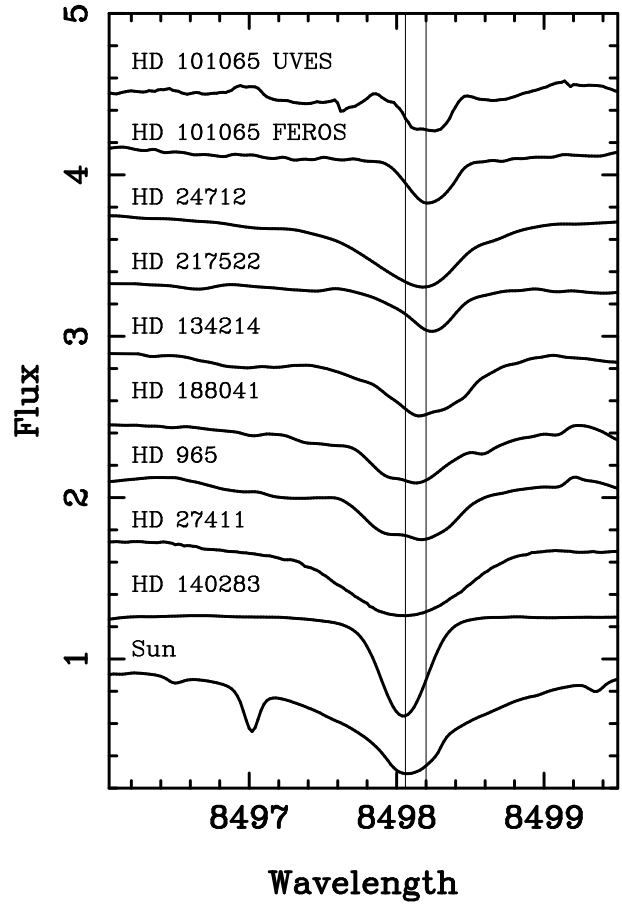}  
      \caption{Spectra of 10 stars in the region of \ion{Ca}{ii} 
              $\lambda$8498.  Spectra have been displaced upward  
              for purposes of illustration by amounts given in the  
              text.  The upper 7 stars are all magnetic stars.  
              HD\,27411 is an Am. HD\,140283 is a well-studied subdwarf. The thin
              vertical lines are at $\lambda$8498.06 and 8498.20.  
              }  
         \label{fig:8498}  
   \end{figure}

Fig.~\ref{fig:8498} shows ten spectra in the region of the  
$\lambda$8498 line.  The spectra are displaced upward,  
respectively by 0.0, 0.27, 0.84, 1.2, 1.6, 2.0, 2.4, 2.8,  
3.2, and 3.7 in units where the continuum is 1.0.  
The thin, vertical lines are at 8498.06 (the solar position) and  
8498.20 (the wavelength for $^{48}$\ion{Ca}{ii}).
The profiles for HD~188041 and HD~965 have double bottoms.  
We attribute this mostly to the Zeeman effect, and  
the reported wavelengths in Tab.~\ref{tab:wl} are for the  
{\it centroids} of these lines.  It does appear that the  
red portion of the profile of both stars is somewhat deeper  
than the violet, and this indicates blending, quite possibly  
by $^{48}$\ion{Ca}{ii}. 
  
%
%

\section{Discussion}  
   
Isotopic anomalies have not yet become established in  
the magnetic sequence of CP stars.  Although they are well known  
to show lines of Hg and Pt, the spectra are more complex than  
those of HgMn stars, so that subtle wavelength  
shifts are easily attributed to blends.  This confusion is  
less likely with lines of the \ion{Ca}{ii} infrared triplet because 
they are located in a region where blending is less severe,  
and they are intrinsically strong.  
  
Abundance anomalies in CP stars are generally attributed to  
chemical fractionations.  The papers cited earlier show that  
this interpretation is not without difficulties.  The $^{48}$Ca  
isotope might be explained in some HgMn stars where there is  
a slight abundance excess of calcium.  In this case, the lighter  
isotopes might be assumed to be pushed out of the photosphere.  
The magnetic stars studied here, are cool, comparable  
in temperature to Am stars, where calcium sinks.  It  
is unclear what kind of fractionation scenario might account  
for an excess of the heavy isotope of calcium in these stars. 
 
Is there a connection between the $^{48}$Ca-stars 
investigated here, and the HgMn objects found by Castelli \& 
Hubrig to show the same anomaly?  The only obvious similarity 
is that both kinds of stars show the most extreme abundances,
and, especially isotopic, anomalies.

Nuclear scenarios were once considered in connection with CP  
stars, but have been widely abandoned.  The $^{48}$Ca isotope  
poses a problem for any nucleosynthetic scheme, as discussed  
for example by Clayton (2003).

\begin{acknowledgements}

We thank the ESO, the ESO Paranal Science Operation Team and NOAO staff for the  
public data archives.  
We acknowledge useful discussions 
with D. J. Bord, W. M. Martin, and H. Stroke.
\end{acknowledgements}  

\end{document}